%% file: ms.tex
\documentclass[10pt,conference]{IEEEtran}
\IEEEoverridecommandlockouts
\usepackage{cite}
\usepackage{amsmath,amssymb,amsfonts}
\usepackage{algorithmic}
\usepackage{graphicx}
\usepackage{textcomp}
\usepackage{xcolor}
\usepackage{array}
\usepackage{multirow}
\usepackage{hyperref}
\usepackage{caption}
\def\BibTeX{{\rm B\kern-.05em{\sc i\kern-.025em b}\kern-.08em
    T\kern-.1667em\lower.7ex\hbox{E}\kern-.125emX}}
\begin{document}

\title{Where Should I Deploy My Contracts? \\
A Practical Experience Report}

\author{\IEEEauthorblockN{C\u{a}t\u{a}lina Laz\u{a}r}
\IEEEauthorblockA{
\textit{Alexandru Ioan Cuza University}\\
Ia\c{s}i, Romania \\
lazarcatalina377@gmail.com}
\and
\IEEEauthorblockN{Gabriela Secrieru}
\IEEEauthorblockA{
\textit{Alexandru Ioan Cuza University}\\
Ia\c{s}i, Romania \\
gabriela.neagu20@gmail.com}
\and
\IEEEauthorblockN{Emanuel Onica}
\IEEEauthorblockA{
\textit{Alexandru Ioan Cuza University}\\
Ia\c{s}i, Romania \\
emanuel.onica@uaic.ro}
}

\maketitle

\begin{abstract}
Blockchain networks provide a reliable trust anchor to decentralized applications (DApps) backed by smart contracts.
The Ethereum ecosystem now encompasses most blockchain networks that provide compatible support for smart contracts code. 
Recently, many Ethereum Layer~2~(L2) rollup solutions emerged, meant to scale the base Layer~1~(L1) network, consequently decreasing transaction fees and diversifying the usage scenarios. 
Furthermore, the number of blockchain providers that offer access to the network infrastructure for both L1 and L2 continuously increases. 
A developer is faced with a multitude of deployment options and must weigh between the gains in costs and the losses in trust that are still an issue with L2. 
A decisive factor in this trade-off can be the use case itself, depending on its security requirements.
Still, the evaluation of costs and performance cannot be ignored and should rely on a set of measurable metrics, although choosing the right metrics can be complicated.
In this practical experience report, we explore the relevance of several such metrics in choosing between different providers and rollups. 
For this purpose, we perform evaluations for two use cases of DApps: a voting DApp with high security demands, suited for L1 deployment, and a cost-sensitive supply chain DApp, where L2 can be an option. 
We analyze a set of basic metrics by comparing these between two highly used access providers, Alchemy and Infura, for the L1 deployment case, and between two of the most popular rollups, Arbitrum One and OP Mainnet (Optimism), for the L2 deployment scenario.
\end{abstract}


\input{intro}

\input{apps}
\input{providers}
\input{rollups}

\input{conclusion}

\section*{Acknowledgment}
We acknowledge the valuable feedback provided by our anonymous reviewers, as well as by Etienne Rivière and Andrei Arusoaie.
This work was supported by a grant from the Romanian Ministry of Research, Innovation and Digitization, CNCS/CCCDI - UEFISCDI, project number 86/2025 ERANET-CHISTERA-IV-SCEAL, within PNCDI IV.

{
\bibliographystyle{IEEEtran}
\bibliography{biblio}
}

\end{document}

%% file: intro.tex
\section{Introduction}
\label{sec:introduction}

Blockchain networks are a disruptive technology that enabled a new source of trust for web applications: an immutable ledger.
This ledger is formed of securely chained blocks, including transaction information submitted by the users of the blockchain platform. 
Ethereum was the first blockchain to support Turing complete smart contracts, small programs that the blockchain nodes can execute. 
A transaction can trigger functions in the contracts that modify the contracts' state.
Changes can be verified because transactions are indexed in the ledger. 
Its immutability is ensured by a consensus protocol executed by network nodes. 

A decentralized application (DApp) is typically designed as a web application, where the backend logic uses smart contracts deployed on a blockchain network.
Multiple DApp use cases can benefit from the decentralized trust anchor that Ethereum provides~\cite{DApps}. 
Still, the current widespread use remains mostly confined to the areas of decentralized finance and gaming~\cite{Bartl23}. 
Arguably, we believe two major factors cause this limitation. 
First, the transaction fees prevent onboarding of users who do not possess the necessary cryptocurrency to pay these fees~\cite{Chandra22}.
Second, functional DApp requirements on security and performance are criteria that make it difficult to choose where to deploy a DApp's smart contracts.

The issue of prohibitive transaction fees has been addressed during last years by multiple approaches. 
A prominent one is scaling the processing power of the main Ethereum network (L1) through rollups, a type of second layer network (L2)~\cite{rollups}.
The rollup nodes can execute transactions as L1 nodes do.
Further, the compressed transaction data, grouped in batches, and the resulting contract states are committed to L1 for trust anchoring.
Rollups come in two flavours: optimistic rollups, which consider transactions implicitly valid and set a challenging time window for proving otherwise, and zk-rollups, which use zero-knowledge proofs for immediate validation.
Delegating transaction execution to either type of rollups has a direct impact on lowering transaction fees. 

However, this solution indirectly augments the second issue above. 
Both rollup types rely on a centralized sequencer to order transactions and post them to Layer 1, raising potential censorship issues~\cite{Koegl23}. 
Moreover, many optimistic rollups don't have a fully reliable proof system yet, which is required in the challenge phase~\cite{Kiepuszewski24}. 
Such aspects still justify executing transactions on L1 for certain use cases, where security concerns might outweigh cost benefits.
In our work, we explore the deployment conundrum a DApp developer is faced with, with a side focus on expanding the use cases outside the typical realm of financial and gaming applications. 

Besides the high security demands of the use case, the limited choice of using L1 may be justified by the expected infrequent user interaction with the DApp.
The access frequency criteria seem to be validated by the current slow shift of gaming DApps to L2, indicating casual use. 
This contrasts with DeFi, where users typically engage for the long term. 
DappRadar~\cite{DAppRadar}, a major DApps directory currently lists 817 game DApps using Ethereum main network and only 229 with deployment on the first three most used rollups.
In comparison, for the DeFi area, 760 DApps are listed for Ethereum L1 and 463 on the respective rollups.\footnote{We note that some listed DApps are deployed on both L1 and L2, and also that statistics include older DApps that haven't shown recent interaction. This justifies the significantly higher number of L1 deployments.}

When limiting the contracts deployment to L1, a difference can still be made by the performance metrics of the provider service chosen for accessing the network infrastructure. 
Currently, there are more than 100 providers offering access to the Ethereum main network, and also for various rollup networks~\cite{CompareNodes}. 
For DApps where L2 is considered for deployment, the difference is made especially by the reduction in transaction costs when using different rollups. 
Currently, L2Beat~\cite{L2Beat}, an L2 monitoring directory, lists over 50 rollups and their number is increasing. 

In this practical experience report, we explore the relevance of a set of metrics in making a deployment choice. 
We consider two use case examples. 
One is for a voting scenario, where security requirements are stringent, but one user is not expected to access the DApp frequently. 
For this, we focus on L1 deployment, where we compare some basic metrics between two popular providers, Alchemy~\cite{Alchemy} and Infura~\cite{Infura}.
The second use case is for a supply-chain scenario, where a user will interact with the DApp more often.
Therefore, we examine L2 deployment for cost reduction, and we compare several related metrics between two of the most used rollup networks, Arbitrum One~\cite{Arbitrum} and OP Mainnet also referred as Optimism~\cite{Optimism}.
We consider this a first step towards a set of deployment metrics that a DApp developer should take into account for a deployment choice. 

Our paper is structured as follows. 
Section~\ref{sec:apps} summarizes relevant technical background information on Ethereum transaction costs and some details about the two DApp use cases we consider. 
Section~\ref{sec:providers} presents our providers comparison for the L1 deployment scenario. 
In Section~\ref{sec:rollups} we discuss the metrics considered in the L2 deployment scenario, using rollups.
Finally, we conclude in Section~\ref{sec:conclusion}.

%% file: apps.tex
\section{Background and Evaluated Applications}
\label{sec:apps}

Our interest is in the decentralized component of a DApp, namely the smart contracts, deployed on either the L1 Ethereum main network or on a L2 rollup network.
We start by presenting some aspects concerning the transaction flow towards the contracts and incurred costs, which are relevant for deciding on evaluation metrics. 
Afterwards, we continue with a brief description of the DApp use cases we consider. 

\subsection{Accessing the Blockchain Network}

The interaction with a network in the Ethereum ecosystem is done through a JSON-RPC protocol.
The software clients used for running a network node implement this, offering an interaction API.
The Ethereum specifications define a baseline JSON-RPC API for the main L1 network~\cite{JSON-RPC}. 
The various clients, either for L1 or L2 networks, can differ in their implemented API, offering sometimes extensions, but normally maintain compatibility with the defined specifications of core functionalities like sending a transaction. 

Irrespective of the L1 or L2 network, a typical access option used by a DApp in a commercial scenario is contracting a provider service that offers network access and guarantees certain QoS parameters. 
A provider runs an infrastructure of nodes, usually for multiple networks, exposing RPC endpoints to customers. 
Providers offer differently priced access plans according to the desired request throughput or the number of simultaneously served DApps. 
Different providers can also offer specific RPC APIs, in addition to the typical ones exposed by the network nodes, to help in the development process. 
While sometimes alternatives can be used by resorting to faster accessible historical data sources for queries, RPC requests are necessary for submitting transactions. 
Therefore, choosing a certain provider can make a difference when considering the same network for deployment. 
In our work, we focus on analyzing how relevant some basic network access metrics are when choosing between providers, such as RPC request latency and failure rate, as we will discuss in Section~\ref{sec:providers}.

\subsection{Transaction Costs}

In the Ethereum ecosystem, a transaction can have the purpose of transferring cryptocurrency, creating a smart contract, or invoking a function in a smart contract that results in contract state changes. 
A transaction has a cost, quantified in units of~\emph{gas}, which can be split into the cost of submitting a transaction and the execution cost. 
This prevents network abuse, such as DoS attempts by transaction flooding or invoking computationally intensive contract code that can incapacitate a node executing it. 

On the main L1 Ethereum network, computing the gas cost of a transaction starts from a fixed amount depending on the transaction's purpose. 
This is incremented per each byte in the encoded transaction, with non-zero bytes adding a higher cost. 
The core component of the client implementations executing the transactions on the network nodes is the Ethereum Virtual Machine (EVM). 
The smart contract functions invoked by transactions are typically written in a high-level language (e.g., Solidity). 
These are compiled into a low-level bytecode, which is executed by the EVM. 
Each low-level instruction adds an execution gas cost defined in the Ethereum specifications. 

The transaction initiator is expected to pay a transaction fee, expressed in Ether (ETH), the Ethereum cryptocurrency. 
For typical Ethereum transactions, this fee is computed by multiplying the total transaction gas cost by the price per unit of gas. 
This price includes a base fee that changes with the network load and a priority fee to incentivize block validators to include the transaction in a block. 

On L2 networks, the transaction fee includes a component like on L1, which accounts for the gas execution cost multiplied by the gas price. 
However, the gas unit price on L2 is typically considerably lower than on L1. 
This execution fee on L2 is summed with a fee for posting the transaction data to L1. 
The calculation method for the L1 posting fee differs across various rollup networks.  
Starting with the adoption of EIP-4844~\cite{EIP4844}, a new transaction type has been introduced, particularly useful for rollups, which permits posting a larger amount of data to L1 in blobs carried along with the transaction.
The blobs are persisted for a limited amount of time. 
This permits lowering the L1 posting fees by introducing a separate \emph{blob gas} cost and a corresponding fee per blob unit.
The L1 data posting process, including the middleware contracts involved, the posting fee computation, and the possible use of EIP-4844 transactions in the flow, is transparent from the DApp perspective.
The transaction initiator, no matter if the contract resides on L1 or L2, must set a single gas cost upper limit parameter willing to cover, and the maximum gas price is willing to pay. 
However, the L2-to-L1 data posting variations between different rollups can influence the total transaction gas cost and the paid fee, as we investigate in Section~\ref{sec:rollups}. 

\subsection{DApp Use Cases}

We consider two separate DApp use cases\footnote{The smart contracts corresponding to the two tested use cases are available at \url{https://github.com/eonica/Smart-contracts-for-DApp-deployment-study}.} to illustrate when semantic factors still make L1 deployment a viable option, although L2 is generally more cost-effective.

The first use case is a voting DApp dedicated to an election process. 
We consider a very simple structure of the application with only two user roles, voter and administrator, and one lightweight smart contract. 
The role of the smart contract is to complement a backend database, used for securely storing vote information, by providing an additional immutability proof. 
The smart contract provides two state changing functions that can be triggered by transactions. 
One is a \emph{commit} function to register the vote hash on the ledger. 
The second is a \emph{reveal} function, triggered by an administrator when retrieving the votes information from the database. 
This function verifies if the vote information in the database corresponds to the hashed replica on the ledger, and marks in the contract state that the reveal verification has passed. 
The commit-reveal scheme's mechanics, such as managing secret salt parameters for voters and ensuring indistinguishability of committed hashes, are handled outside the public smart contract logic. 

We can observe that the smart contract logic is minimal in this DApp use case, strictly for strengthening the integrity of the votes stored in the database by backing these with immutable proof. 
A regular DApp user interacts with the contract using just the \emph{commit} function, which should be executed only once per voter during an election. 
We can derive, therefore, a first factor that directly relates to the use case semantics, the \emph{transaction frequency} to the contract code, which in this case is low per user, implying reduced individual financial costs. 
However, its overall aggregate per contract can be high during time intervals corresponding to election terms, when large numbers of voters may trigger transactions. 

Another factor is the \emph{transaction flow security}.
This is valid for any DApp, but some use case semantics increase its importance.
For elections, it can be critical to finalize within limited time intervals, without missing votes that can change results. 
Therefore, the probability of transaction delay or censorship should be reduced to a minimum.
This situation cannot be excluded with centralized components in the transaction path, like the sequencer in L2 rollups.
In conjunction with the transaction frequency per voter, we consider these solid criteria for L1 deployment in this use case. 

The second use case we consider is of a supply chain management DApp for the agriculture sector, specifically for vegetable markets. 
This use case leverages the blockchain immutability for trusted tracking of the production and operational flow. 
A client can trace a batch of goods and identify all parties involved from seed acquisition to final storage and display.
The use case involves multiple actor roles: seeds and pesticide sellers, farmers, distributors, warehouses and stores, and a DApp administrator. 
Commercial actors must register with the DApp by paying a fee and getting approved by the administrator through two specific transactions.
The rest of the smart contract logic is kept simple to lower costs.
Other transactions mostly register simple events that track the status of product batches at various stages in the supply chain. 

Most transactions that change a batch status and register events (\emph{buySeeds}, \emph{germinateSeeds}, \emph{plant}, \emph{stimulate}, \emph{harvest}) are initiated by the farmer.
A single farmer is expected to submit such transactions for multiple batches of goods, having frequent interaction with the DApp. 
Other actors, like distributors and stores, are linked with a lower number of possible transactions they can trigger (\emph{transport}, \emph{store} and respectively \emph{display}).
However, their frequency is again expected to be high, being called for multiple product batches originating from multiple farmers. 
Therefore, this DApp will have an overall high transaction frequency per user. 

While transaction flow security is still important, delaying or censoring transactions is expected to have a somewhat lesser impact in the context of this use case semantics.
Missing transactions do not mandatorily prevent finalizing the actual supply-chain flow for a batch of goods. 
A post-verification observing a breaking point in the status changing transactions can actually indicate the specific actor or step where a problem occurred, which is the purpose of a supply-chain tracking implementation.
We can conclude that this use case would be a good candidate for L2 deployment. 

We further evaluated each of the two DApps using two L1 providers for the voting use case and respectively two L2 rollups for the supply-chain use case, to determine the relevance of specific performance and cost metrics.
Our evaluations were conducted from April to June 2024. 
We discuss the results we obtained in the section that follows.

%% file: providers.tex
\section{Providers Metrics Comparison}
\label{sec:providers}

If the main Ethereum network (L1) is deemed viable for smart contracts deployment according to use case semantics as discussed in Section~\ref{sec:apps}, DApp developers still have a choice to make between providers.
The principal metrics we consider in the provider decision are the \emph{latency} and \emph{throughput} of RPC requests corresponding to submitted transactions. 
We investigate their relevance for two high-profile providers, Alchemy~\cite{Alchemy} and Infura~\cite{Infura}.

Our evaluation setup is as follows. 
We deployed the voting DApp smart contract on the Ethereum Sepolia test network, a public replica of the main network that uses the same consensus and gas pricing protocols.
We deployed the rest of our DApp implementation (i.e., the frontend code and the backend database storing the extended vote information) on the Microsoft Azure cloud platform, in 5 different geographical zones: East US, Canada Central, West Europe, Brazil South and Japan West.
Each virtual machine instance was allocated 2 CPUs from a 64-core AMD EPYC 7763 processor, clocked at 2.45 GHz and running 2 threads per core. 
We dispatched batches of POST HTTP requests from a Microsoft Azure DevOps agent located in the East US region, which simulates the user side of the DApp.
These were routed through the DApp backend, triggering transactions towards providers for committing hashes of the database stored vote information. 
When transactions are confirmed or an error is received, an HTTP response is sent back to the user side. 

\begin{table}[t]
  \caption{Summary of results for provider non-stress tests \\ (RT: Response Time, FR: Fail Rate)}
  \label{tab:nonstress}
\begin{tabular}{| m{0.1\linewidth} | m{0.12\linewidth} || m{0.12\linewidth} | m{0.12\linewidth} | m{0.12\linewidth} | m{0.12\linewidth} |} 
 \hline
 Region & Provider & Batch 50/60 RT (ms) & Batch 100/100 RT (ms) & Batch 50/60 FR (\%) & Batch 100/100 FR (\%) \\ 
 \hline\hline
East & Alchemy & 398.6 & 383.2  & 0 & 0 \\\cline{2-6}
US & Infura & 225.98  & 305.44 & 0 & 0 \\\hline
Canada & Alchemy & 485.62  & 472.08  & 0 & 0 \\\cline{2-6}
Central & Infura & 377.84  & 449.94  & 0 & 0 \\\hline
West & Alchemy & 1153.26  & 1233.73  & 6 & 5 \\\cline{2-6}
Europe & Infura & 1026.26  & 922.08  & 0 & 4 \\\hline
Brazil & Alchemy & 1507.32  & 1375.51  & 6 & 8 \\\cline{2-6}
South & Infura & 1254.44  & 1238.33  & 2 & 4 \\\hline
Japan & Alchemy & 2004.16  & 1802.43  & 18 & 50 \\\cline{2-6}
West & Infura & 1772.7  & 1723.35  & 10 & 44 \\\hline
\end{tabular}
\vspace{-10pt}
\end{table}

In our voting scenario, one user is expected to send just one vote per election term. 
Still, the DApp is supposed to handle a larger amount of requests for votes generated by multiple users. 
We executed a first set of two non-stress tests, within the throughput limits offered by the free provider plans by sending 50 requests over 60 seconds (Batch 50/60), and respectively 100 requests over 100 seconds (Batch 100/100).
We measured the average response time on the user side and verified the error rate for the considered throughput.  
The results are summarized in Table~\ref{tab:nonstress}.

The average response time difference between the two providers is 178.35 ms for Batch 50/60 and 125.56 ms for Batch 100/100.
This reveals the providers' added RPC latency impact, which seemed to favor Infura at the time of our tests. 
This difference metric becomes less relevant as the distance between the user's site (East US) and the DApp deployment site increases.
For the Canada Central deployment, the impact on the response time when choosing a different provider is still higher than the delay added by the different deployment sites. 
However, when the DApp deployment is moved outside the North American continent the geographical distance inflicts an added delay in the response time that ranges from 4x to more than 10x the time difference between providers. 

Providers restrict the throughput of RPC requests, such as transactions and queries, based on the chosen pricing plan. 
However, there are differences in how providers quantify throughput.
Some use a metric that accounts for computation costs based on the type of request. 
Others use a simple limit set in requests per second.
At the time of our tests, Alchemy's throughput was measured in compute units (CU), as mentioned in the first case.  
The free plan used in our evaluation was limited to 330 CU per second.  
A raw transaction for a vote commit in our DApp was charged 250 CU, limiting throughput to 1.32 transactions per second. 
Infura offered a free plan that allowed 10 requests per second, equating to 5 transactions per second, since each transaction requires an additional request.\footnote{Since our evaluation, Alchemy switched to quantifying throughput in requests per second, maintaining a monthly CU cap, while Infura now uses a custom metric of credits per second that covers computation costs. 
}

Despite keeping the sending rate within the throughput limits of the providers, we could observe a fail rate that increases as the distance between the user and the DApp deployment site grows. 
All errors resulted from the interaction with the providers. 
The vast majority were reported to have the reason of \emph{replacement fee too low}. 
Each Ethereum transaction includes a nonce, counting the transactions emitted by a certain account. 
Our use case considers one transaction per account, but splitting the funds for transaction fees over as many accounts as transactions were sent was cumbersome. 
Therefore, an account was used multiple times as a transaction initiator. 
This should not be a concern for our tests because it doesn't affect the measurements from the two providers differently.
However, before sending a new transaction, an automatic RPC request is sent to retrieve the latest nonce for the initiator account, counting pending transactions too.
Alchemy did not charge this in CU, and for Infura this results in two requests per transaction, still below the throughput limit.
The issue is that as the distance increases, the answer for this request can arrive just before earlier transactions reach a pending state, likely due to the added network delays. 
Consequently, the new nonce will not be increased. 
A transaction with the same nonce from the same account must have a higher price to replace a pending transaction, otherwise resulting in an error as reported.

We note again that this is not an issue which would normally happen in the voting scenario, where each account would send only one transaction. 
However, it reveals an interesting fact for a scenario where the same account would submit multiple transactions within a short time frame. 
The error occurred because we did not vary the price in submitted transactions.
Otherwise, if a new transaction is issued at a higher price, a previous pending transaction may be silently dropped and replaced without the user's knowledge.
This situation must be handled by the DApp implementation.
Some wallets prevent this by requiring final confirmation before allowing new transactions from the same user. 

The second error we encountered had a much lower frequency (maximum 2\%) and as the reported reason \emph{could not coalesce}, being specific to Alchemy. 
This referred to the situation when the provider denied the request due to exceeding the throughput limit.
We believe this to be caused by the same factor when network delays due to the distance result in requests reaching the provider at a different rate than the one used when sending. 
Given that Alchemy's free plan throughput limit was very low, it is expected that this could be easily bypassed in such situations.
However, Alchemy's documentation mentions a certain tolerance of CUs throughput limit in the context of their elastic scaling~\cite{AlchemyCU}. 
Therefore, we decided to conduct two additional stress tests: sending 50 requests over 20 seconds (Batch 50/20) and 100 requests over 50 seconds (Batch 100/50).
These exceed Alchemy's throughput limit by about 1 transaction per second but remain within Infura's limit.
We summarize the results in Table~\ref{tab:stress}.

\begin{table}[t]
  \caption{Summary of results for provider stress tests \\ (RT: Response Time, FR: Fail Rate)}
  \label{tab:stress}
\begin{tabular}{| m{0.1\linewidth} | m{0.12\linewidth} || m{0.12\linewidth} | m{0.12\linewidth} | m{0.12\linewidth} | m{0.12\linewidth} |} 
 \hline
 Region & Provider & Batch 50/20 RT (ms) & Batch 100/50 RT (ms) & Batch 50/20 FR (\%) & Batch 100/50 FR (\%) \\ 
 \hline\hline
East & Alchemy & 473.2 & 386.92  & 52 & 38 \\\cline{2-6}
US & Infura & 330.46  & 301.44 & 6 & 1 \\\hline
Canada & Alchemy & 524.14  & 567.19  & 58 & 36 \\\cline{2-6}
Central & Infura & 399.96  & 347.25  & 8 & 1 \\\hline
West & Alchemy & 1160  & 1093.67  & 66 & 64 \\\cline{2-6}
Europe & Infura & 940.28  & 958.15  & 58 & 50 \\\hline
Brazil & Alchemy & 1474.9  & 1338.08  & 70 & 65 \\\cline{2-6}
South & Infura & 1459.58  & 1274.04  & 54 & 53 \\\hline
Japan & Alchemy & 2115.84  & 1780.1  & 72 & 68 \\\cline{2-6}
West & Infura & 1787  & 1753.91  & 68 & 66 \\\hline
\end{tabular}
\vspace{-10pt}
\end{table}

The average response time difference between the two providers stayed consistent with earlier tests, showing 166.16 ms for Batch 50/20 and 126.23 ms for Batch 100/50, along with delays from the distance to the deployment site.  
Unsurprisingly, the failure rate for Alchemy increased due to bypassing the throughput limit. 
The corresponding \emph{could not coalesce} error reason was dominant in all tests for Alchemy. 
Interestingly, the percentage of errors citing~\emph{replacement fee too low} rose sharply, increasing more than tenfold in some deployments, while the throughput rate only increased 2-3 times. 
All errors reported by Infura are due to this reason.
We can observe an increase for Alchemy too for this error, sometimes with more than 20\%, although this seems limited by the throughput limit failure, which we assume takes precedence in reporting. 
This suggests that raising the throughput, although even within the plan limits (Infura's case), might create some backpressure in the transaction requests processing, preventing these from reaching the pending state fast enough. 
We acknowledge that this is an assumption that would require further investigation.

%% file: rollups.tex
\section{Rollups Metrics Comparison}
\label{sec:rollups}

The choice to deploy DApp smart contracts on a specific rollup (L2) will likely be driven by cost savings.  
This might be intricate to estimate due to the varied implementations of different rollups.
We analyze the relevance of several metrics concerning this aspect on two popular rollups, Arbitrum One~\cite{Arbitrum} and OP Mainnet also referred to as Optimism~\cite{Optimism}.
We start with the \emph{gas price estimation} provided by the rollup networks. 
Afterwards, we look at the split between the \emph{gas used on L2} for the transaction execution and \emph{gas used on L1} for transaction data posting.
Further, we observe how these are reflected in the total~\emph{transaction fee}.
We complete our investigation with the comparison between \emph{transaction soft confirmation latencies}, which can supplement cost metrics in deciding towards a rollup. 

We deploy our supply-chain DApp smart contracts to the Arbitrum Sepolia and Optimism Sepolia networks, which are test replicas of the main rollup networks, as in the L1 case. 
The provider is supposed to offer similar access conditions to both rollups and should not have any impact on the comparison of the above metrics. 
We used Alchemy in our tests due to its richer integrated developer support. 
Most of the functions in our supply-chain use case perform a simple status setting and log an event, resulting in close costs for the corresponding transactions. 
We submit batches of 100 transactions for each function. 
Due to space limits, we will focus on the \emph{buySeeds} function for comparison, as it has a slightly heavier implementation.
The corresponding transaction information will be compressed and posted to L1. 
This includes the function call with two parameters: the 20-byte long address of a seller and a variable sized string identifying the product (we used a 6-byte string in our tests). 
The function code is guarded by two modifiers, wrappers that verify the roles of the caller account and the seller.
The code logic starts by initializing the goods batch in the supply chain, then logs an event, and finally increments the batch counter.

We can obtain several metrics from the transaction receipts provided by both rollups for each transaction.  
A typical Ethereum transaction receipt includes a standard set of fields~\cite{receipt}. 
We are particularly interested in two metrics: \emph{effectiveGasPrice}, which indicates the price per gas unit at the time of execution when the sender's balance is deducted, and \emph{gasUsed}, which totals the gas units consumed by the transaction. 
These fields do not have identical semantics in the receipts returned by the two rollups.
This comes as a result of how the rollups compute the charges for the separate L2 execution and L1 posting.
On Arbitrum, the \emph{effectiveGasPrice} applies to both layers, while on Optimism, it only applies to L2 execution. 
Similarly, on Arbitrum, the \emph{gasUsed} field counts the total gas spent by the transaction, using a conversion to L2 gas units as common measurement unit~\cite{arbirtrumfees}, while in Optimism counts only the gas spent for L2 execution~\cite{optimismcosts}. 

The first metric we analyze is the gas price on the L2 rollups. 
A DApp typically needs to estimate this by querying the network before a transaction is submitted. 
This provides the users with some idea about the total fee for executing the transaction and allows them to agree with it or adjust or adjust the limits on priority and base fees per gas unit.
A difference between the two analyzed rollups is that Arbitrum does not charge priority fees while Optimism does. 
Gas prices can fluctuate between the estimated amount and the actual price shown on transaction receipts.  
For our batch of 100 \emph{buySeeds} transactions, the gas price on Arbitrum Sepolia remained set at the fixed floor of 0.01 Gwei\footnote{1 Ether (ETH) = $10^9$ Gwei = $10^{18}$ Wei} both in the initial estimation and the effective confirmed price, with zero variation between transactions.
We consider this stability to be a direct consequence of the lack of priority fees. 
On Optimism, the initial gas price estimation varied within the range of just 5 Wei across transactions. 
However, we noticed a difference between the paid price and the initial estimation, having an average of 200051 Wei.
This led to a paid average of 0.0012 Gwei per gas unit. 
Given the two observed situations, we consider the difference between confirmed and estimated gas prices to be a relevant metric concerning stability.
This is particularly significant for small batches, where it is not expected to observe high price variations. 

Rollups sometimes include different additional information in transaction receipts, which can be helpful in a cost analysis. 
Arbitrum adds a~\emph{gasUsedForL1} field.
This does not represent the effective gas units for L1 posting, but rather a normalized value based on the variable L1 gas price. 
In brief, it can be considered that the value of this field represents a conversion from the total price charged for L1 gas to L2 gas units.
This allows interpreting the \emph{effectiveGasPrice} in the receipts as a uniform price rate per gas unit for both L1 and L2~\cite{arbirtrum2Dfees}. 
Optimism adds specific fields to the transaction receipt, which quantify the actual L1 payment: \emph{l1GasUsed} counting the effective gas units spent for the L1 transaction data communication, \emph{l1GasPrice} representing the price paid per L1 gas unit, \emph{l1BlobBaseFee} indicating the price charged for blob storage, two scalar values used in computing a weighted common gas price multiplier for the normal gas and blob gas, and \emph{l1Fee} that provides the resulting total amount paid as L1 charges.

The gas spent on L2 can be determined on Arbitrum by subtracting \emph{gasUsedForL1} from the total \emph{gasUsed} specified by the transaction receipts, and on Optimism directly by checking the \emph{gasUsed} field.
As we expect the gas spent on L2 for the transaction execution is consistent between the two rollups (82534 gas units for the \emph{buySeeds} transaction). 
Both Arbitrum and Optimism are EVM compatible and the evaluation confirms that the compiled smart contract function results in identical costs. 
However, we note that small differences can appear between different EVM based rollups, especially when these require the use of modified compilers. 
In general, we consider this metric to be of low relevance in a comparison. 

\begin{figure}[t]
  \centering
  \includegraphics[width=\linewidth]{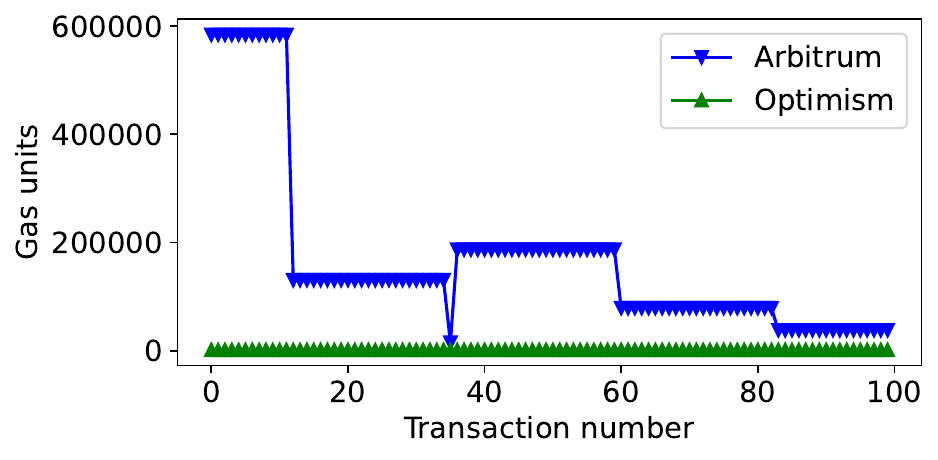}
  \caption{\label{gasL1}L1 quantified gas used for the buySeeds transactions 
  }
  \vspace{-10pt}
\end{figure}

For the gas spent on L1, we refer strictly to the information provided in the transaction receipts.
The results are depicted in Figure~\ref{gasL1}.\footnote{At the time of our evaluation, both rollups were using blobs for L1 data posting. Optimism was running the Ecotone upgrade. Since then, the transaction compression algorithm has been improved in the Fjord upgrade.} 
Normally the gas amount for identical transactions should be a constant value.
This is reflected by the Optimism measurement that reports this metric fixed at 1600 units for L1. 
As discussed before, what Arbitrum reports as L1 gas units is basically a conversion into L2 gas, which accounts for gas price variations on L1. 
Therefore, this metric unusually varies in this case, having an average value of 169467 gas units.
From an effective payment perspective, the L1 gas is consistently higher priced than L2 gas, making this a relevant metric in a comparison. 
However, as we observe, its reported value is highly dependent on the rollup mechanics.
In this particular case, since the L1 gas of Arbitrum does not represent effectively L1 gas units, it is difficult to draw a conclusion without comparing the final fees.

\begin{figure}[b]
  \vspace{-10pt}
  \centering
  \includegraphics[width=\linewidth]{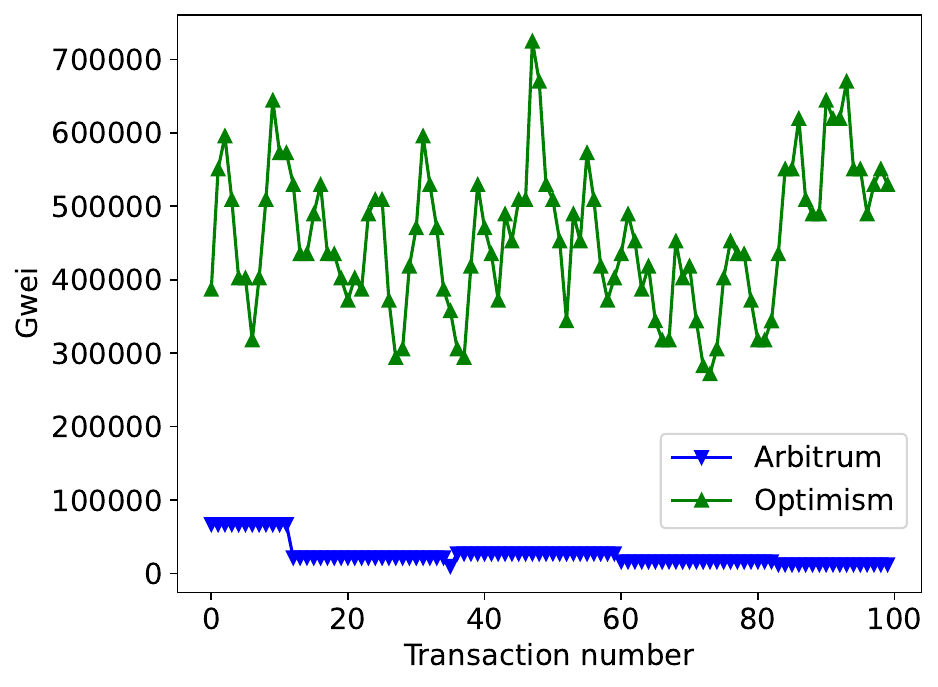}
  \caption{\label{priceTX}Total fee of the buySeeds transactions}
  \vspace{-5pt}
\end{figure}

In Figure~\ref{priceTX} we depict the total transaction fees on the two rollups.
These confirm the influence of the L1 price in the converted L1 gas variation on Arbitrum.
The total fee follows the same precise pattern, with a resulting average of 25196 Gwei per transaction.
Despite the apparently lower L1 gas amounts and the lower price per L2 gas unit, Optimism fees seem to be influenced more by the initially uncaught L1 gas price variations, resulting in higher fees than Arbitrum, with an average of 457632 Gwei per transaction. 
The total fee appears to remain a considerably more robust comparison criterion than its components discussed before.
This is somewhat expected since the total confirmed fee is the final post-transaction metric.
What our analysis confirms is that a developer should not rely solely on the gas amounts, which can also be estimated pre-transaction. 
The gas price, in particular on L1, can have a significant impact on the final fees. 

\begin{figure}[b]
  \vspace{-10pt}
  \centering
  \includegraphics[width=\linewidth]{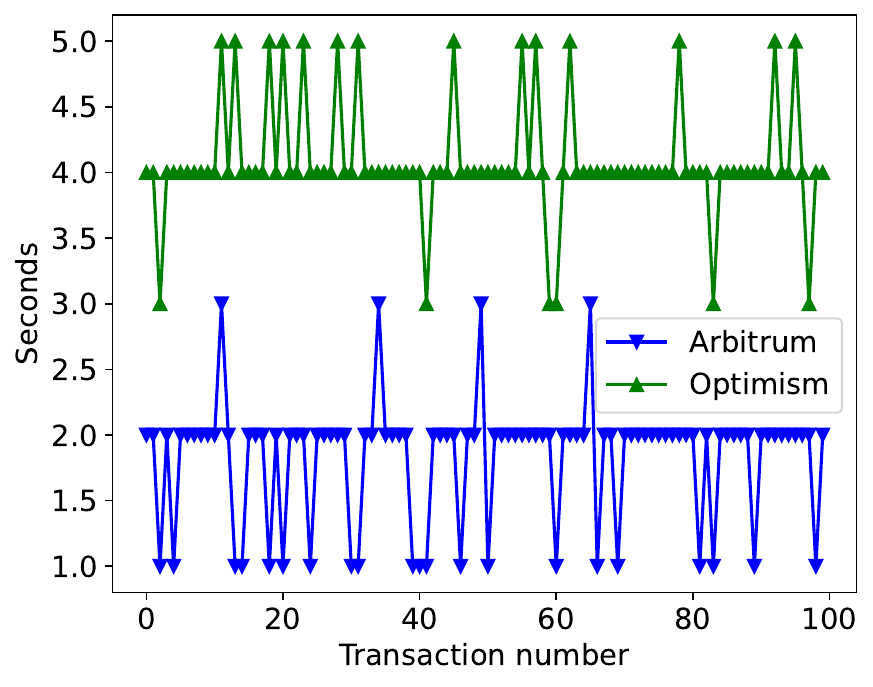}
  \caption{\label{confirmation}Soft confirmation latency for the buySeeds transactions}
  \vspace{-5pt}
\end{figure}

We finally consider a performance metric, the soft confirmation latency at the sequencer side. 
This measures the time in seconds from transaction creation to when the sequencer issues the receipt.
We illustrate the results in Figure~\ref{confirmation}.
The average duration for Arbitrum Sepolia is 1.83 seconds, while for Optimism Sepolia is 4.08 seconds.
Interestingly, the worst times for Arbitrum are in the same range as the best times for Optimism. 
We can speculate that this indirectly correlates with the previous observations concerning the cost metrics that suggest a higher load on Optimism.
However, confirming this requires further investigation.

%% file: conclusion.tex
\section{Conclusion}
\label{sec:conclusion}

In this practical experience report, we investigated the relevance of several metrics in selecting a medium for deploying the smart contracts of a DApp.
We argued that certain semantic factors of the use case might provide a first indication either towards L1 or L2.
In the situation of L1 deployment, the choice is between different providers.
We compared Alchemy and Infura.
From our analysis, we conclude that the RPC request latency might be a relevant metric for choosing a provider for smart contracts deployment if the end users of the DApp are not far from the DApp backend deployment site. 
Otherwise, the request latency importance can diminish considerably.
On the other hand, an analysis of the RPC throughput corresponding to transactions can be relevant even with fixed throughput limits because it might reveal issues that relate to network delays.  

The findings we observed for the L1 deployment scenario also apply to L2 deployments, which similarly involve choosing a provider to access a specific rollup network.
However, in the L2 deployment case, we preferred to focus on analyzing cost metrics that might be relevant in choosing between different rollups.
We note that the results we discussed depend heavily on network dynamics and do not serve as a long-term indicator for favoring a certain rollup between the two we compared, Arbitrum and Optimism.
As a general finding, these showed that the L1 gas reported by different rollups is not a robust metric for comparison unless the rollups use the same computation base.
Also, our investigation suggests that the L2 execution costs have less impact than the L1 posting charges, which would be typically influenced by the network load and might arguably correlate with the soft confirmation latencies.

Our study does not aim to cover all deployment solutions available or provide an exhaustive list of metrics for comparison. 
We acknowledge that other relevant metrics exist, in particular for deciding between rollups, such as average unused block space that can impact transaction throughput, or available liquidity that can correlate to immediate user capacity to access DApps. 
Our measurements are a first step towards a more advanced monitoring solution that could offer developers detailed information for making deployment decisions. 
We plan to extend this study by evaluating other providers, rollups and metrics, which should consolidate our findings.